\documentclass{iau} 
\usepackage{graphicx}
\usepackage{hyperref}

\def\apj{ApJ}%

\title[La Serena School for Data Science.] 
{La Serena School for Data Science: multidisciplinary hands-on education in the era of big data.}

\author[A. Bayo, et al.]   
{A. Bayo $^{1,2}$
  M. J. Graham$^3$, D. Norman$^4$, M. Cerda$^{5,6}$, G. Damke$^{7,8}$, A. Zenteno$^8$, \and C. Ibarlucea.$^8$}

\affiliation{$^1$Inst. de F\'isica y Astronom\'ia, Universidad de Valpara\'iso, Chile \\ email: {\tt amelia.bayo@uv.cl} 
$^2$N\'ucleo Milenio de Formaci\'on Planetaria (NPF)
$^3$ California Institute of Technology.
$^4$NSF's OIR Lab, Tucson, AZ
$^5$Inst. of Biomedical Sciences \& Center for Medical Informatics and Telemedicine, Universidad de Chile.
$^6$ Biomedical Neuroscience Institute, Santiago, Chile.
$^7$Instituto de Investigaci\'on Multidisciplinar en Ciencia y Tecnolog\'ia, Universidad de La Serena
$^8$Association of Universities for Research in Astronomy (AURA).
}

\pubyear{2021}
\volume{367}  
\jname{Education and Heritage in the Era of Big Data in Astronomy}
\editors{A.C. Editor, B.D. Editor \& C.E. Editor, eds.}
\begin{document}

\maketitle

\begin{abstract}

La Serena School for Data Science is a multidisciplinary program with six editions so far and a constant format: during 10-14 days, a group of $\sim$30 students (15 from the US, 15 from Chile and 1-3 from Caribbean countries) and $\sim$9 faculty gather in La Serena (Chile) to complete an intensive program in Data Science with emphasis in applications to astronomy and bio-sciences.

The students attend theoretical and hands-on sessions, and, since early on, they work in multidisciplinary groups with their ``mentors" (from the faculty) on real data science problems.
The SOC and LOC of the school have developed student selection guidelines to maximize diversity. 

The program is very successful as proven by the high over-subscription rate (factor 5-8) and the plethora of positive testimony, not only from alumni, but also from current and former faculty that keep in contact with them.

\keywords{astroinformatics, statistics, data bases, surveys, machine learning, big data.}
\end{abstract}

\firstsection 
\vspace{-0.1cm}
\section{Introduction: the challenge}
The volume and complexity of astronomical data continue to grow as the current generation of surveys come online (Gaia, SDSS / APOGEE, etc).  Beyond these challenges, astronomers will need to work with giga-, tera-, and even peta-bytes of data in real time in the era of LSST. Large data-sets pose the challenge of developing and using new tools for data discovery, interoperability and access, and analysis. 

This framework brings also new opportunities for interdisciplinary research in applied mathematics, statistics, machine learning, and other areas under active development. Astronomy provides a sandbox where scientists can come together from diverse fields to address common challenges within the ``Big Data" paradigm.  

But of course, astronomy is not alone. Society’s inexorable digitization of data and the rapidly evolving Internet are driving the need for global transformation of data intensive science in many fields. Indeed, ``Big Data” now impacts nearly every aspect of our modern society, including retail, manufacturing, financial services, communications \& mobile services, health care, life sciences, engineering, natural sciences, art \& humanities.  

Clearly, our research leadership hangs on whether the next generation can be productive within the petabyte-sized data volumes generated in different domains. Unfortunately, the development of ``Big Data" and ``Artificial Intelligence" (AI) related skills (including, for example machine learning) is not present commonly enough in the University curricula worldwide. This is particularly true in Chile (with a few examples that have emerged in the last years like the astroinformatics initiative from Universidad de Chile), and also in the US outside of the main / top Universities. 

For instance, reports in Chile yield a deficit in highly trained AI related professionals in the thousands per year (as claimed in the ``Pol\'itica Nacional de Inteligencia Artificial" draft presented by the Chilean government in December 2020).

La Serena School for Data Science (\href{http://www.aura-o.aura-astronomy.org/winter\_school/}{LSSDS}) emerged in 2013 with the leadership of AURA, aiming at covering part of the gap in training via a combine effort of Chilean and US individuals and institutions.

\vspace{-0.5cm}
\section{A diverse school with a rich history}

LSSDS targeted since the beginning students either in their last years of undergraduate school, or the first years of graduate school (with a new pilot program involving high-school students). The school has welcomed students with majors or minors in either mathematics, statistics, physics, computer sciences, astronomy, and more recently, bio-related subjects.

The program is very intense, and lasts between 10 and 14 days with constant interactions between the core faculty and the students. The teaching philosophy is project oriented with $\sim$34\% of the time spent on lectures (covering basic to intermediate level statistics, basics of Data Science, Machine Learning, Distributed computing and data-bases), $\sim$ 23\% of the time spent on hands-on labs (that settle the content of the lectures), and the remaining $\sim$43\% of the time is devoted to work on the group's project with one of the faculty acting as the mentor. The transition between the three types of activities are gradual through the school, with the first days being more ``lecture" heavy, and the last days of the school being focused solely on project time (\cite{Cabrera17}). 

The school also offers opportunities for social interactions, with the ``top" activity being the visit to several of the AURA telescopes stargazing close to the summit.

The groups for the projects, with typically four students, are purposely design to maximise the diversity among the students. The projects themselves are proposed by the core faculty and tend to have different steps in difficulty with a possible open-ended final goal. Some of these projects have in fact resulted in successful observing proposals, continued collaboration, and even the replication of parts of the school by some of the students in their home institutions. 

Diversity in the school is pursued since the very beginning, starting with the student selection process. When sorting the students applications, we try to keep a balance between the different minors and majors, we look for diversity of institutions (keeping a balance between students from big / well-known schools and those less renowned), gender balance, and nationality.

We believe that most of the success of the program can be grouped in two aspects: the inherent richness obtained from the multidisciplinary and diversity conscious student (and faculty) selection process, and the strong commitment from the faculty. Regarding the latter, more than half of the professors stay through the whole school serving as lecturers, hands-on instructors, and as mentors for the student group projects.

Another very relevant factor is the very strong commitment and support from AURA, NOIRLab, the LSST Corporation, NSF, REUNA, CONICYT, CORFO and other Chilean institutions that, since the very beginning understood the need to train the new generation of, not only astronomers, but scientists, in data driven problems, which particularly benefit from diversity and multidisciplinarity.

\end{document}